# Vacuum as a hyperbolic metamaterial

Igor I. Smolyaninov

*Department of Electrical and Computer Engineering, University of Maryland, College Park, MD 20742, USA*

**As demonstrated by Chernodub, vacuum in a strong magnetic field behaves as a periodic Abrikosov vortex lattice in a type-II superconductor. We investigate electromagnetic behavior of vacuum in this state. Since superconductivity is realized along the axis of magnetic field only, strong anisotropy of the vacuum dielectric tensor is observed. The diagonal components of the tensor are positive in the *x* and *y* directions perpendicular to the magnetic field, and negative in the *z* direction along the field. As a result, vacuum behaves as a hyperbolic metamaterial medium. If the magnetic field is constant, low frequency extraordinary photons experience this medium as a (3+1) Minkowski spacetime in which the role of time is played by the spatial *z* coordinate. Spatial variations of the magnetic field curve this effective spacetime, and may lead to formation of "event horizons", which are analogous to electromagnetic black holes in hyperbolic metamaterials. We also note that hyperbolic metamaterials behave as diffractionless "perfect lenses". Since large enough magnetic fields probably had arisen in the course of evolution of early Universe, the demonstrated hyperbolic behavior of early vacuum may have imprints in the large scale structure of the present-day Universe.**

Physical vacuum is a complicated medium which is often understood as a "boiling soup" of virtual particles, such as photons, electron-positron pairs, quarks and gluons. Under some circumstances these virtual particles may become real, such as in the

Shwinger effect [1], in which strong enough electric field in vacuum creates real electron-positron pairs. Another such effect has been discovered very recently by Chernodub [2]. It appears that a strong magnetic field forces vacuum to develop real condensates of electrically charged $\rho$ mesons, which form an anisotropic inhomogeneous superconducting state similar to Abrikosov vortex lattice [3] in a type-II superconductor. This effect follows from the consideration of motion of a free relativistic spin $s$ particle in an external magnetic field $B$. The energy levels of the particle are [4]

$$E_{n,s_z}^2 = m_\rho^2 + p_z^2 + (2n - 2s_z + 1)|eB| \qquad (1)$$

where $n \geq 0$ is integer, and $s_z$=-$s$, …, $s$ is the spin projection on the field axis. The ground state energy of the $s$=1 charged $\rho$ mesons is thus

$$m_\rho^2(B) = m_\rho^2 - eB \qquad (2)$$

which indicates that at large enough magnetic fields

$$B > B_c = \frac{m_\rho^2}{e} \approx 10^{16} T \qquad (3)$$

vacuum becomes unstable. It spontaneously generates positively and negatively charged $\rho$ meson condensates. Chernodub also demonstrated that these condensates are superconducting and spatially inhomogeneous. They form a periodic Abrikosov lattice (Fig.1b) of superconducting vortices separated by dielectric gaps of the order of $L_B = \sqrt{2\pi/|eB|}$. As a result, vacuum has no conductivity in the $x$ and $y$ directions perpendicular to the magnetic field, while it behaves as a superconductor along the $z$ axis [4]. In this paper we investigate electromagnetic behavior of vacuum in the high magnetic field state. Since superconductivity is realized along the axis of magnetic field only, strong anisotropy of the vacuum dielectric tensor is observed. The diagonal components of the tensor are positive in the x and y directions perpendicular to the magnetic field, and negative in the z direction along the field, which means that vacuum behaves as a hyperbolic metamaterial medium. As demonstrated in [5,6], electrodynamics of such media is quite unusual. It may be briefly summarized as follows.

Let us start with a non-magnetic ($\mu = 1$) uniaxial anisotropic material with dielectric permittivities $\varepsilon_x = \varepsilon_y = \varepsilon_1$ and $\varepsilon_z = \varepsilon_2$ and assume that dispersion of the dielectric permittivity can be neglected. The wave equation in such a material can be written (see for example [7]) as

$$-\frac{\partial^2 \vec{E}}{c^2 \partial t^2} = \vec{\vec{\varepsilon}}^{-1} \vec{\nabla} \times \vec{\nabla} \times \vec{E} \qquad (4)$$

where $\vec{\vec{\varepsilon}}^{-1}$ is the inverse dielectric permittivity tensor calculated at the center frequency of the signal bandwidth. Any electromagnetic field propagating in this uniaxial material can be expressed as a sum of the "ordinary" and "extraordinary" contributions, each of these being a sum of an arbitrary number of plane waves polarized in the "ordinary" ($\vec{E}$ perpendicular to the optical axis) and "extraordinary" (vector $\vec{E}$ parallel to the plane defined by the k–vector of the wave and the optical axis) directions. We will be interested in the behavior of the extraordinary portion of the field. Let us define our "scalar" extraordinary wave function as $\varphi = E_z$ (so that the ordinary portion of the electromagnetic field does not contribute to $\varphi$). Equation (4) then yields:

$$\frac{\partial^2 \varphi}{c^2 \partial t^2} = \frac{\partial^2 \varphi}{\varepsilon_1 \partial z^2} + \frac{1}{\varepsilon_2}\left(\frac{\partial^2 \varphi}{\partial x^2} + \frac{\partial^2 \varphi}{\partial y^2}\right) \qquad (5)$$

While in ordinary crystalline anisotropic media both $\varepsilon_1$ and $\varepsilon_2$ are positive, this is not necessarily the case in metamaterials. In the hyperbolic metamaterials (Fig.1a) considered for example in [8,9] $\varepsilon_1$ and $\varepsilon_2$ have opposite signs. In the visible frequency range these metamaterials are typically composed of multilayer metal-dielectric or metal wire array structures [9]. Optical properties of such metamaterials are quite unusual. For example, it was demonstrated theoretically in [8,10,11] that there is no usual diffraction limit in a hyperbolic metamaterial. This prediction has been confirmed experimentally in [12,13]. In the absence of dispersion, eq.(5) in the case of $\varepsilon_1 > 0$ and $\varepsilon_2 < 0$ looks like the Klein-Gordon equation for a massless field in a flat (3+1) Minkowski space-time:

$$\frac{\partial^2 \varphi}{\varepsilon_1 \partial z^2} = \frac{\partial^2 \varphi}{c^2 \partial t^2} + \frac{1}{(-\varepsilon_2)}\left(\frac{\partial^2 \varphi}{\partial x^2} + \frac{\partial^2 \varphi}{\partial y^2}\right) \qquad (6)$$

However, it is the $z$-coordinate which assumes the role of a time-like variable in this equation.

Now let us apply the standard metamaterial description to vacuum subjected to high magnetic field $B > B_c$. Diagonal components of the vacuum dielectric constant may be obtained using Maxwell-Garnett approximation [9]:

$$\varepsilon_2 = \varepsilon_z = \alpha \varepsilon_s + (1-\alpha)\varepsilon_d, \quad \varepsilon_1 = \varepsilon_{x,y} = \frac{2\alpha \varepsilon_s \varepsilon_d + (1-\alpha)\varepsilon_d(\varepsilon_d + \varepsilon_s)}{(1-\alpha)(\varepsilon_d + \varepsilon_s) + 2\alpha \varepsilon_d} \qquad (7)$$

where $\alpha$ is the volume fraction of the superconducting phase, and $\varepsilon_s < 0$ and $\varepsilon_d > 0$ are the dielectric permittivities of the superconducting and dielectric phase, respectively. In the low frequency limit $-\varepsilon_s >> \varepsilon_d = 1$. Therefore, eq.(7) may be simplified as follows:

$$\varepsilon_2 = \varepsilon_z \approx \alpha \varepsilon_s < 0, \quad \varepsilon_1 = \varepsilon_{x,y} = \frac{1+\alpha}{1-\alpha} > 0 \qquad (8)$$

The $\rho$ meson condensate density $\alpha$ has been calculated in [2,4] as

$$\alpha(B) = C_\phi^2 \frac{m_q^2(B)}{G_V^2}\left(1 - \frac{B_c}{B}\right), \tag{9}$$

where $C_\phi$=0.51 is a constant, $m_q(B)$ is the quark mass, and $G_V$ is the vector coupling of four-quark interactions. The dielectric constant of an ideal superconductor for frequencies well below the superconductor's gap frequency is given by the zero-loss Drude model [14]:

$$\varepsilon_s = 1 - \frac{\omega_s^2}{\omega^2}, \tag{10}$$

where $\omega_s = c/\lambda_L$, and $\lambda_L$ is the London penetration depth. In the type-II superconductors the zero-loss approximation is not perfectly valid due to the following effect. When a low frequency ac field is applied to the Abrikosov vortex lattice, electric currents flowing through the vortices lead to the Lorentz force acting between the vortices. As a result, vortices move with a drift velocity $v_d$ proportional to the current density $j$, which leads to a very small non-zero dc resistance of the unpinned type-II superconductor [15]. Since no pinning centers may exist in vacuum, the Abrikosov lattice of charged $\rho$ meson condensates must be in the "vortex liquid" state, and the non-zero dc resistance of vacuum in the $z$-direction is defined by the viscosity of this liquid. Theoretical estimate of the vortex liquid viscosity has been performed by Bardeen and Stephen [15]. However, application of this theory to the $\rho$ meson vortex fluid is beyond the scope of this paper. Observational effect of the vortex fluid viscosity consists in cutting off low frequency singularity in eq.(10). As a result, low frequency behavior of the extraordinary optical field in vacuum is well described by eq.(6) with finite negative $\varepsilon_2$. Therefore, low frequency extraordinary photons perceive this medium as a flat (3+1) effective Minkowski spacetime. However, it is the $z$-coordinate which assumes the role of a time-like variable in this effective spacetime. As has been demonstrated recently [16], spatial variations of the effective $\varepsilon_1$ and $\varepsilon_2$ permittivities of the hyperbolic medium may give rise to appearance of the electromagnetic black holes, as perceived by extraordinary photons. The high magnetic field vacuum state may experience similar behavior if magnetic field changes its magnitude as a function of spatial location. For example, if the magnetic field is critical $B=B_c$, an effective metric signature transition is observed [5,17,18] which inevitably leads to appearance of effective horizons [16].

Finally, let us note that hyperbolic metamaterials behave as diffractionless "perfect lenses" [10-13]. Due to their hyperbolic dispersion law

$$\frac{\omega^2}{c^2} = \frac{k_z^2}{\varepsilon_1} + \frac{k_x^2 + k_y^2}{\varepsilon_2} \qquad (11)$$

the absolute value of the extraordinary photon $k$ vector is limited only by the metamaterial structure parameter. In the case of vacuum subjected to high magnetic field, the $k$ vector is limited by the Abrikosov lattice periodicity $k<K_{max}=2\pi/L_B$. As noted by Chernodub [4], large enough magnetic fields probably had existed in the very early Universe [19]. Therefore, the demonstrated hyperbolic behavior of early vacuum may have imprints in the large scale structure of the present-day Universe.

**Figure Captions**

**Figure 1.** (a) Schematic view of the "wired" hyperbolic metamaterial geometry: an array of aligned conductive wires is embedded into a dielectric host. (b) According to Chernodub [2], a similar structure spontaneously develops in vacuum when it is subjected to high magnetic field. Notice that the Meisner effect is absent due to anisotropy of vacuum superconductivity. (c) Hyperbolic dispersion relation of extraordinary photons propagating inside a hyperbolic metamaterial is illustrated as a surface of constant frequency in k-space. When $z$ coordinate is "timelike", $k_z$ behaves as effective "energy".

10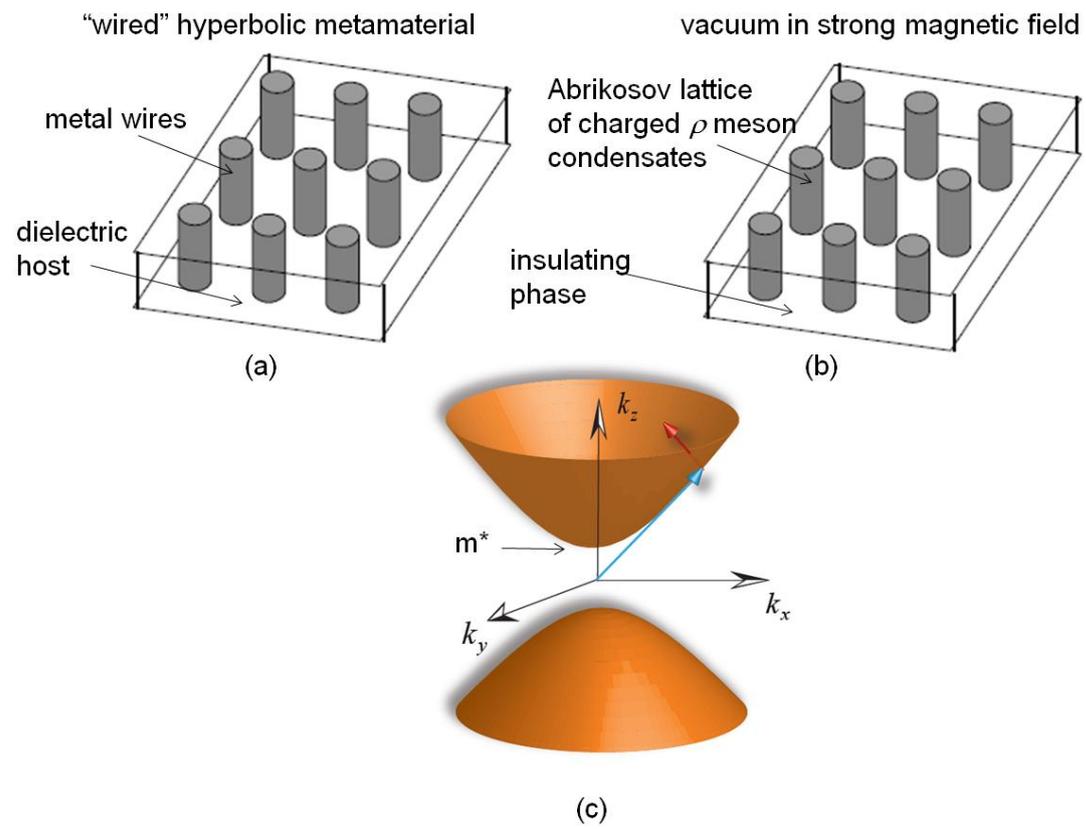

Fig. 1